\documentclass[sn-mathphys,Numbered]{sn-jnl}

\usepackage{pdflscape}
\usepackage{graphicx}
\usepackage{multirow}
\usepackage{amsmath,amssymb,amsfonts}
\usepackage{amsthm}
\usepackage{mathrsfs}
\usepackage[title]{appendix}
\usepackage{textcomp}
\usepackage{manyfoot}
\usepackage{booktabs}
\usepackage{algorithm}
\usepackage{algorithmicx}
\usepackage{algpseudocode}
\usepackage{listings}
\usepackage{booktabs}
\usepackage[table,xcdraw]{xcolor}

\usepackage{soul} 
\newcommand{\lh}{\textcolor{black}}
\begin{document}

\title{Synchronization between media followers and political supporters during an election process: towards a real time study}

\author[1]{\fnm{Rémi} \sur{Perrier}}\email{remi.perrier@cyu.fr}

\author*[1]{\fnm{Laura} \sur{Hernández}}\email{laura.hernandez@cyu.fr}

\author[2,3]{\fnm{J. Ignacio} \sur{Alvarez-Hamelin}}\email{ihameli@fi.uba.ar}

\author[2,3,4]{\fnm{Mariano G.} \sur{Beiró}}\email{mbeiro@udesa.edu.ar}

\author[1]{\fnm{Dimitris} \sur{ Kotzinos}}\email{dimitrios.kotzinos@cyu.fr}

\affil[1]{\orgdiv{Laboratoire de Physique Théorique et Modélisation, UMR-8089 CNRS}, \orgname{CY Cergy Paris Université}, \orgaddress{\street{2, Av.Adolphe Chauvin}, \postcode{95300}, \state{Paris}, \country{France}}}

\affil[2]{\orgdiv{Universidad de Buenos Aires}, \orgname{Facultad de Ingeniería}, \orgaddress{\street{Paseo Colón 850}, \city{Buenos Aires}, \postcode{ACV1063C}, \country{Argentina}}}

\affil[3]{\orgdiv{CONICET,  Universidad  de  Buenos  Aires}, \orgname{INTECIN},
\country{Argentina}}

\affil[4]{ \orgname{Universidad de San Andrés}, \orgaddress{\street{Vito Dumas 284}, \city{Victoria}, \postcode{B1644BID}, \country{Argentina}}}

\abstract{We present an   analysis of the dynamics of  discussions in  Twitter (before it became X) among supporters of various candidates in the 2022 French presidential election, and followers of different types of media. Our study demonstrates that we can automatically detect the synchronization of interest among different groups around specific topics at particular times.

We introduce two complementary methods for constructing dynamic semantic networks, each with its own advantages. The \textit{growing aggregated network} helps identify the reactivation of past topics, while the \textit{rolling window network} is more sensitive to emerging discussions that, despite their significance, may appear suddenly and have a short lifespan. These two approaches offer distinct perspectives on the discussion landscape. Rather than choosing between them, we advocate for using both, as their comparison provides valuable insights at a relatively low computational and storage cost. 
Our findings confirm and quantify, on a larger scale and in an automatic, agnostic manner, observations previously made using more qualitative methods.
We believed this work represents a step forward in developing methodologies to assess equity in information treatment, an obligation imposed by law on broadcasters that use broadcast spectrum frequencies in certain countries.}


\maketitle

\section{Introduction}\label{sec1}

The explosive growth of the number of empirical studies about the evolution of social opinion is fueled by the widespread usage of social networks and by the fast development of the techniques that allow the collection and analysis of the large amount of generated data.
Although some of these works deal with particular case studies, others focus on the development of improved or radically new techniques to make sense of the huge data sets that are nowadays accessible~\cite{9259192}. 

 In this work, we study the evolution of the discussion in Twitter during a presidential campaign, taking into consideration the interactions between different actors in the network.
 This type of study attracts the interest of two different but overlapping communities of researchers: Social Network Analysis and Computational Social Science, each one endowed with some specific characteristics, mainly related to their history. While the former is rooted in the works of J.Moreno~\cite{moreno_who_1934}, who introduced the importance of interpersonal links in sociology long before online social networks could even be imagined, the latter, as it name acknowledges, emerged as a consequence of the need of turning the massive amount of data generated by digital media into sensible material. 

Specifically, in opinion dynamics studies one is interested in tracing the temporal evolution of the opinion of the social actors around one or several topics that are discussed in society. There are different ways to collect online data in order to understand opinion evolution: targeting selected users and collecting all the messages they have posted in some online platform, collecting all the messages that contain some selected words in the text --considered as keywords for the particular subject of interest--   using tagged messages, etc. These techniques involve a strong intervention of the researcher from the start, thus assuming a deep knowledge of the case study that helps to decide who are the important actors and why, to define an ontology, etc. Nevertheless, when performed with care, many of such studies successfully showed the emergence of interesting patterns. The existence of polarization or ``lateralization" of the actors in the platform~\cite{gaumont2018reconstruction,rusche2022}, the correlation between some opinions about non-political subjects and political affiliation~\cite{chavalarias2023new} and the existence of filter bubbles~\cite{grossetti2021reducing} or echo chambers~\cite{cinelli2021echo}, or the interplay between cognitive biases and platforms' algorithms in the diffusion of information~\cite{chavalarias2023toxic} are some examples.

An alternative and more \textit{agnostic} procedure, which does not require external information to select some discussion topics a priori, is to let them emerge from the raw data. This approach has been successfully applied to describe the political landscape during electoral campaigns~\cite{mussi_reyero_evolution_2021} or to study the interaction between the topics treated in traditional media and the online discussions of their followers during the COVID19 pandemic~\cite{schawe_understanding_2023}. 

Most of these works extract the topics under discussion by aggregating the collected data over the whole studied period, and then they trace the evolution of the discussions among the  platforms' users around these topics. However, recent  studies targeted  the evolution of topics themselves, using dynamical topic tracking techniques. Some of them are based on LDA  and work either by  computing topics in different time windows, tracking their evolution~\cite{zhang2019tracking,malik2013topicflow} or using a modified version, called SeqLDA \cite{blei_latent_2003, bogdanowicz2022dynamic}. Recent alternatives to these LDA based methods have been proposed like the combination of a label propagation algorithm with cascade information diffusion model~\cite{sattari2018cascade} or in a different and more application-oriented approach, an elaborated framework for social network analysis combining semantic annotation, Linked Open Data, semantic search, and dynamic result aggregation components~\cite{maynard2017framework}. 

If one wants a way to perform a first analysis of the data with minimum  external knowledge of the system, an  alternative is to use co-occurrence networks. In particular, studying the structure of the co-occurrence networks of tags chosen by the users, like \textit{hashtags} in Twitter (now X platform), one can get an insight of the evolution of the topics under discussion. In Ref.~\cite{lorenz2018tracking} communities detected in networks of different time windows are related using a continuous time random walk.

Here, elaborating on our previous works~\cite{mussi_reyero_evolution_2021, schawe_understanding_2023}, we aim at performing a dynamical comparative study between the discussions held by supporters of different political groups and those held by users who prefer different media, during the  campaing for the presidential elections in France in 2022. At a difference with the previously cited works, here we change the paradigm and we use dynamical networks instead of static ones that are aggregated over the whole period, in an attempt to perform a \textit{real time}  approach to the problem.

A key point to consider when pushing these studies beyond the aggregated network framework, is to account for the effect of  memory in social networks.  We propose two different scenarii for the collection and treatment of data: one involving all the previous posts in the platform since the beginning of the data collection process, and another one with a bounded memory. We discuss the consequences of using each of them. Finally we compare the results with those obtained by following the static network paradigm, which leads to a static pool of topics.

\section{Methods}\label{sec2}

\subsection{Data acquisition and pre-processing}
\label{data_acquisition}

\lh{Our study is based on data captured from the Twitter platform (before it became X) during the last presidential campaign in France, from September $1^ {\text{st}}$ 2021 to June $1^ {\text{st}}$ 2022. We have collected all the activity (tweets and retweets) published by the \textit{active} followers of the political candidates of the French presidential election in 2022 or several French media of different kind (journals, radio, continuous news TV chains, online media). We consider as \textit{active followers} those who tweeted or retweeted anything during the period.} Our selection of French news outlets comprises media across all the ideological spectrum\footnote{News Media and Political Attitudes in France, \url{https://www.pewresearch.org/journalism/fact-sheet/news-media-and-political-attitudes-in-france/}, Retrieved 02 March 2025}.

\lh{In order to mitigate organized external influence and astroturfing~\cite{chavalarias2023toxic}, we limited our sample to users  who declared a location in any region, department, or city of France, as well as the country itself. These users are labelled as \textit{``in France''}.} 

\lh{We define as \textit{supporters} of a given candidate those followers whose retweets of the candidate's posts constituted at least 75\% of their total retweets of any candidate. This notion is extended to the followers of media, and in this case we say that the users thus defined have a \textit{preferred media}.}

\begin{figure}[t]
    \centering
    \includegraphics[width=\linewidth]{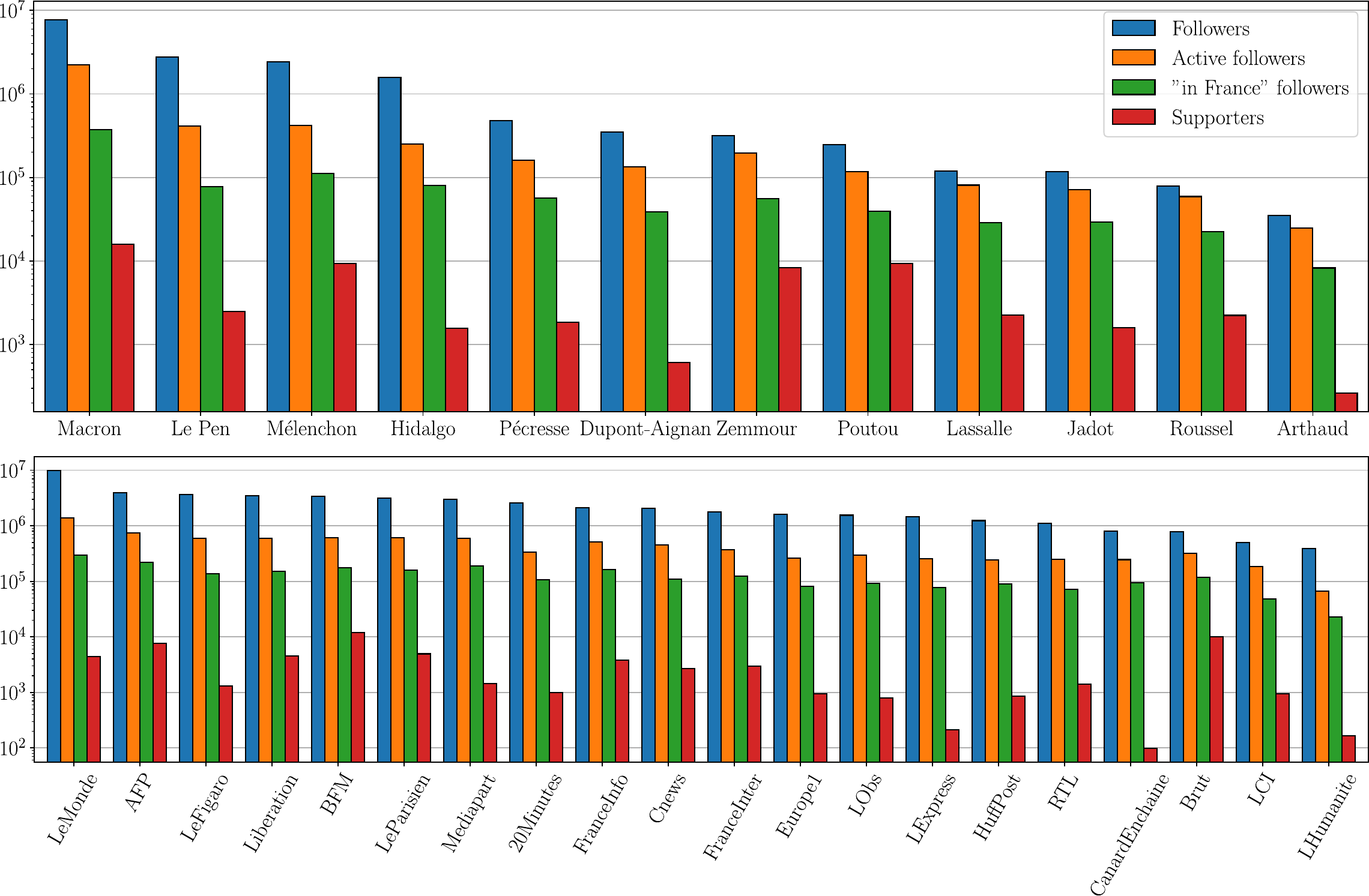}
    \caption{Breakdown of the different types of followers for the official candidates (top) and selected media (bottom).``Followers": total number of accounts following the candidate/media on Twitter as of June 2022.``Active": followers of the candidate/media who publish posts (tweets) or forward (retweet) those of other users. ``in France": Followers with a self-declared location in France.``Supporters": Twitter accounts where more than 75\% of their retweets of candidates/media corresponding to a given candidate/media.}
	\label{stats_candidates_media}
\end{figure}

\lh{Fig.~\ref{stats_candidates_media} illustrates the number of  the different kinds of followers for the considered candidates and media.}

\lh{Following  the hypothesis already used in previous works~\cite{cardoso_topical_2019,mussi_reyero_evolution_2021, schawe_understanding_2023}, which assumes that the hashtags chosen by the users carry a meaning about the subject of the tweet,  we build a \textit{semantic network} where nodes represent the hashtags and a link between a pair of nodes means that the two hashtags have been used in the same tweet. As many different users can choose to put the same pair of hashtags in their tweets (or more, see Fig.~\ref{distrib_hashtags_per_tweet} for the distribution of hashtags per tweet),  the resulting network is non-directed and weighted. However, it should be noted that we increase the weight of the links only once per user,  to diminish the effect of automated accounts. Table~\ref{stats_capture} shows some statistics of the dataset involved in this work.}

\begin{figure}
    \centering
    \includegraphics[width=0.9\linewidth]{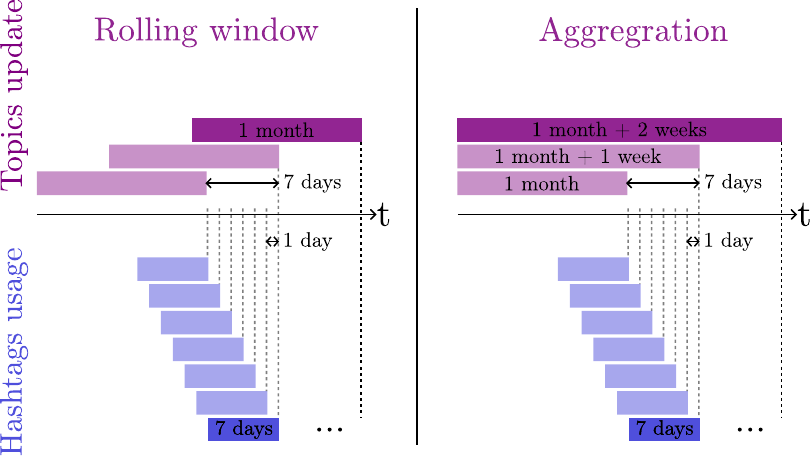}
    \caption{Illustration of the updating scheme for the hashtags co-occurrence graph and the supporters’ hashtags usage}
    \label{update_rolling_aggreg}
\end{figure}

\lh{In order to keep our study as near as possible to a \textit{real-time} analysis, we propose two ways of constructing dynamical semantic networks that allow us to explore the effects of memory in these networks. In both cases, with the intention of having enough data, we start by aggregating a  network during the first month of the studied period. Then for the \textit{rolling network } case one week after the end of the month, the network is shifted in such way that the hashtags used during this week enter in the network and those that have not been used during the last month are ``forgotten" and leave the network, as shown in top left panel of Fig.~\ref{update_rolling_aggreg}. 
Alternatively, in the \textit{ growing aggregated network} case,  the network grows by incorporating the new hashtags that could have appeared in the last week, as shown in the top right panel of Fig.~\ref{update_rolling_aggreg}.  
For comparison, we also use the \textit{static case}, where the network is accumulated over all the period before performing a dynamical analysis of the attention that the users pay to different topics.}

\begin{table}[t]
\begin{tabular}{@{}|l|r|@{}}
\hline
Total number of users                 & 22M             \\ \hline
\rowcolor[HTML]{C5C5C5} 
Number of active users                & 4.5M            \\ \hline
Number of active ``in France" users & 822k            \\ \hline
\rowcolor[HTML]{C5C5C5} 
Total number of tweets                & 1.3B            \\ \hline
Total number of hashtags              & 334M            \\ \hline
\rowcolor[HTML]{C5C5C5} 
Number of co-occurrences              & 84M             \\ \hline
\end{tabular}
\caption{Basic statistics of the dataset.}
\label{stats_capture}
\end{table}

\subsection{Tracking topics}

\lh{As a given topic of discussion may be referred to by different hashtags, finding \textit{communities} in the semantic network -a partition where groups of nodes are significantly more connected among them than with nodes in  other groups- leads to the automatic emergence of  the topics discussed in the platform  from the topological properties of the semantic network~\cite{cardoso_topical_2019}.}

In this paper we used the community detection algorithm OSLOM~\cite{lancichinetti_finding_2011} \lh{which allows for overlapping communities. }
Topic extraction 
 follows the updating scheme previously explained for the two dynamical networks, and is done once for the static semantic network.
 It should be noted that the two methods proposed to construct the dynamic semantic network guarantee that, unlike in the static network approach,  we never use information from the future to determine the topics. In the worst case, we admit a lag of one week in the update  of the daily usage of topics by the different groups of users.
 
\subsection{Studying user's attention dynamics}
\label{similarities}

\lh{In order to study the evolution of the attention that users  pay to the different topics, we proceed as in Refs.~\cite{mussi_reyero_evolution_2021,schawe_understanding_2023} by computing the daily usages  of hashtags corresponding to each topic, for all the users. Specifically, for a given day  $d$ we compute the usages in a window that spans one week in order to reduce well known daily usage fluctuations. At a difference with previous works, here the window is not centered at day $d$, because  to perform a \textit{real time} like study, we need to avoid using information from the future. The procedure is depicted in the bottom of Fig.~\ref{update_rolling_aggreg}, for both methods of update of the semantic networks.} 

\lh{With this data we can build  a daily user-topic matrix  $ U \in N \times N_T$, which records all the times that each candidate uses a hashtag that corresponds to a given topic}. 

\lh{Then, the row vectors $\Vec{u_i}, i \in [1,N] $, of this matrix  belong to the hyper-space of dimension $N_T$,  the number of detected topics, and gives  the number of daily uses of hashtags corresponding to each topic, by user $i$. Accordingly, the vector $\Vec{T}=\sum_i^{N} \Vec{u_i} $  gives the total usage of each topic by all the population.}

We can then define: 
\begin{equation}
    \Vec{v_i}= \frac{\Vec{u_i}}{||\Vec{u_i}||_1} - \frac{\Vec{T}}{||\Vec{T}||_1}  \enspace;
\end{equation}

where $||\cdot||_1$ represents the sums of all components of the vector. 
 Therefore each component of vector $\Vec{v_i}$ informs us about whether  user $i$ has talked, in proportion, about the corresponding topic, more or less than  the global population.

\lh{As the length of the vector will be associated to the activity of the user, we normalize it so as to take into account only the direction that each vector points at  in the topic space.}
\begin{equation}
    \Vec{d_i}= \frac{\Vec{v_i}}{||\Vec{v_i}||_2}  \enspace;
\end{equation}
where $||\cdot||_2$ denotes the standard Euclidean norm, and $d_i$ is the {\em user description vector}. 

\lh{It should be noticed that for the evolving networks the topics are recalculated at each update of the matrix, so that its number, quality, and the corresponding user description vectors depend on the period for which the semantic network holds, and on the method of construction of  such networks (growing aggregated, rolling window, static).}

\lh{Using the description vectors, the cosine similarity between two users is given by their  scalar product: } 
\begin{equation}
    s(i,j) = \langle \Vec{d_i},\Vec{d_j} \rangle \enspace.
\end{equation}

\lh{And extending this definition to groups of users, as shown in Ref.~\cite{mussi_reyero_evolution_2021, schawe_understanding_2023}, 
the \textit{cross-group similarity} is defined as,}
\begin{equation}
    s(G_a,G_b)= \frac{1}{|G_a|\cdot|G_b|} \sum_{i\in G_a, j\in G_b} \langle \Vec{d_i},\Vec{d_j} \rangle = \langle  \Vec{D_{G_{a}}} , \Vec{D_{G_{b}}}  \rangle \enspace.
\end{equation}

with $\Vec{D_G}$ the average description vector of the group of users $G$,  
\begin{equation}
   \Vec{D_G} = \sum_{i\in G} \frac{\Vec{d_i}}{|G|}\enspace.
\end{equation}

Note that when $G_a=G_b$ we obtain the \textit{self-similarity} which is a measurement of cohesion within a group.
It is important to note that \lh{when $G_a \neq G_b$}  then $G_a$ and $G_b$ should be disjoint \lh{to avoid a spurious increase of the similarity lead by common users. While this is by definition the case for groups of supporters of a given party, care must be taken when computing the similarity among supporters of a party and users having a preferred medium.}

\lh{In the following we will study how the different methods of computing the semantic network, with and without long memory, affect the evolution of the similarities between different groups. We will also investigate how the cross-similarity among groups of supporters of different political parties and those of users with a preferential medium evolves during the electoral campaign.}

\section{Results}\label{sec3}

\subsection{General properties of the two evolving semantic networks}

\lh{As described in Sec.\ref{data_acquisition}, we build two different evolving hashtag co-occurrence graphs, one with long term memory and another which only keeps the memory of the last month. }

\lh{In this section we present the evolution of the standard metrics on the resulting non-directed and weighted graphs}~\cite{10.5555/1809753}.
\lh{Specifically, for the clustering coefficient we use the following definition}: 
\begin{equation}
cc=\sum_{\forall i \in V} \frac{cc(i)}{|V|}\enspace,
\end{equation}
\lh{where $V$ denotes the set of vertices, $|V|$ its cardinal, and 
$cc(i)=\sum_{\forall i \neq j \neq k} a_{ij}a_{jk}a_{ki} $. Furthermore  we use the definition of assortative coefficient given by Newman~\cite{newman2002assortative}.}

\begin{figure}[H]
	\centering
  \includegraphics[width=\linewidth]{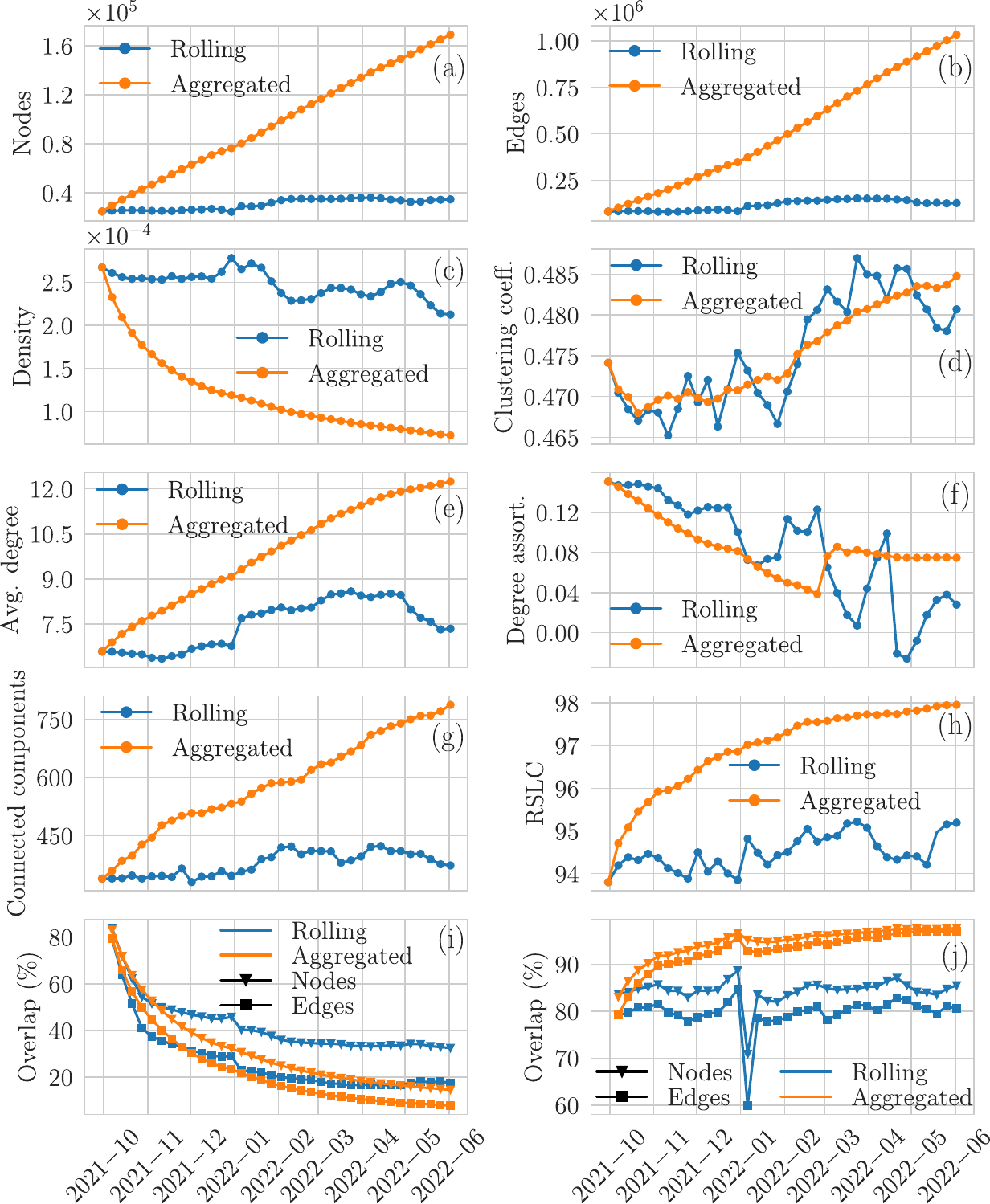}
	\caption{Comparison of the evolution of the global network properties between the rolling window and growing aggregated co-occurrence graphs. The first four rows show standard network metrics: (a) number of nodes, (b) number of edges, (c) density, (d) clustering coefficient, (e) average degree, (f) degree assortativity, (g) number of connected components, (h) relative size of the largest component. The bottom row, shows in (i) the  persistence of the  co-occurrence graphs: fraction of nodes/edges present at the beginning of the capture, that are still present at time $t$ , and in (j) the ``instantaneous" renewal of the  graphs: fraction of nodes/edges present at time $t$ that were already present at time $t-1$.}\label{graphs_properties}
\end{figure}

Figure~\ref{graphs_properties}  
\lh{shows the comparison of the time evolution of the computed metrics of the two  networks.}
Panels (a)-(c) illustrate the expected differences between a network with an average steady number of nodes and a growing one. In particular in  panel (c) the decrease of the density of the growing aggregated graph is just a dilution effect, while the almost constant trend observed for the rolling graph shows that there is a balance between ``new" and ``forgotten" hashtags.

\lh{The average degree (e), number of connected components (g)  and relative size of the largest components (h) show again a trivial increase for the growing aggregated network and seem to stabilize for the rolling one. }
On the other hand, the  \lh{clustering} coefficient \lh{shown in panel (d)} is almost constant with a very slight increase (notice the scale)  in the aggregated case, while  in the rolling case, it fluctuates around this slightly increasing trend. These two apparently contradictory effects --strong decrease of the density and slow increase of the clustering coefficient-- can be understood by examining the behavior of the clustering coefficient and the density with incoming nodes (hashtags). For an incoming hashtag to increase the clustering, it must at least form a new triangle, this means that it must connect with two other hashtags that are connected themselves. This new node will always increase the denominator of the density  $D = \frac {2E}{N (N-1)} $, in $2N$, strongly impacting  the dilution effect although it may form only a few triangles more. The alternative  case where the clustering is increased  by  existing hashtags that become connected by new joint adoptions, does not decrease the density. So the panels (c) and (d) jointly show that there are always new hashtags arriving, which are mainly responsible for the slightly increasing trend in the clustering and compatible with the dilution effect. 
 On the other hand, the fluctuations observed around the increasing trend for the rolling windows method, are  caused by  hashtags that could have formed triangles earlier, but being less used, they leave the network and are  replaced by new ones, therefore provoking a sudden temporary decrease in the clustering coefficient.
The observed slight increase of the clustering in spite of the massive dilution of the network, also reveals that there is a counterbalancing, cohesive effect driven by the semantic coherence of the hashtags. In other words, people introducing new hashtags do not do it at random, on the contrary, for the sake of their diffusion, it is useful to couple them with already trending ones.

The spike observed at the end of February 2024 in the assortativity curve of the aggregated case (panel (f) ), is related to the start Ukrainian war.

\lh{The two panels (i) and (j), at the bottom of Fig.}~\ref{graphs_properties} \lh{show the \textit{retention rate}, the fraction of nodes or edges at day $d$ that were already present at the origin (panel d) and the \textit{instantaneous rate}, the fraction of nodes or edges at time $t$ that were already there at time $t-1$. The former shows a trivial behaviour for the growing aggregated case, again due to dilution; on the other hand, for the rolling windows case, the retention rate seems to stabilize after a transient,   showing a ``core" that is present since the beginning of the capture. For the instantaneous rate, we observe that after a first increase the aggregate case saturates (no new hashtags enter the system, and therefore the new tweets come to reinforce existing links). For the rolling windows case, there are small fluctuations in the  presence of the nodes or the edges, except from a notch that corresponds to the hashtags associated to the end of the year festivities: once they have passed they will massively leave the network.   }

\subsection{Hashtags and topics entropy}

\lh{A global measurement of the distribution of the attention among the different subjects of discussion may be obtained by computing the entropy of the hashtags and of topics usage distributions. The entropy of the hashtags distribution,  is  independent of the community structure, and informs about the diversity of the \textit{vocabulary} used in the platform. On the contrary, as the topics emerge from the community structure, their entropy may depend on the way  the network is updated.}

\lh{Specifically at day $d$ we compute the hashtag usage distribution and we define the hashtag entropy as:} 

\begin{equation}
  S_d=\sum_{h} p_d(h) \log  p_d(h) \enspace, \label{entropy_d}
\end{equation}
 
\noindent where \lh{$p_d (h)$ is the probability distribution calculated as the ratio between the number of unique users that have
used hashtag $h$ on the week preceding day $d$ (included) and the number of different pairs (hashtag, user) within the same time frame. By considering unique users we diminish the influence of very active accounts (e.g., spammers). We calculate the entropy daily with the same 
rolling time frame of seven days to remove the well known influence of the lower activity on weekends.}

For the topic entropy, a similar definition is applied for the two different update methods for the semantic network. As in both cases the update is made every week, we compute the entropy of topics usage during each week: 
\begin{equation}
  S_w=\sum_{t} p_w(t) \log  p_w(t) \enspace, \label{entropy_w}
\end{equation}

\lh{where now $p_w(t)$  is the probability distribution calculated as the ratio between   the number of unique users
that have used topic $t$, and the number of different pairs (topic, user), both computed during the month preceding week $w$.} \lh{By \textit{usage of a topic} we mean the sum of the usages of the hashtags constituting the topic.} The choice of computing the topic usages of the week $w$ by cumulating over the month that precedes and includes $w$  is aimed at avoiding discontinuities at the moment of updating the network (and thus  the topics) which is done weekly. 

\begin{figure}[t!]
	\centering
     \includegraphics[width=\linewidth]{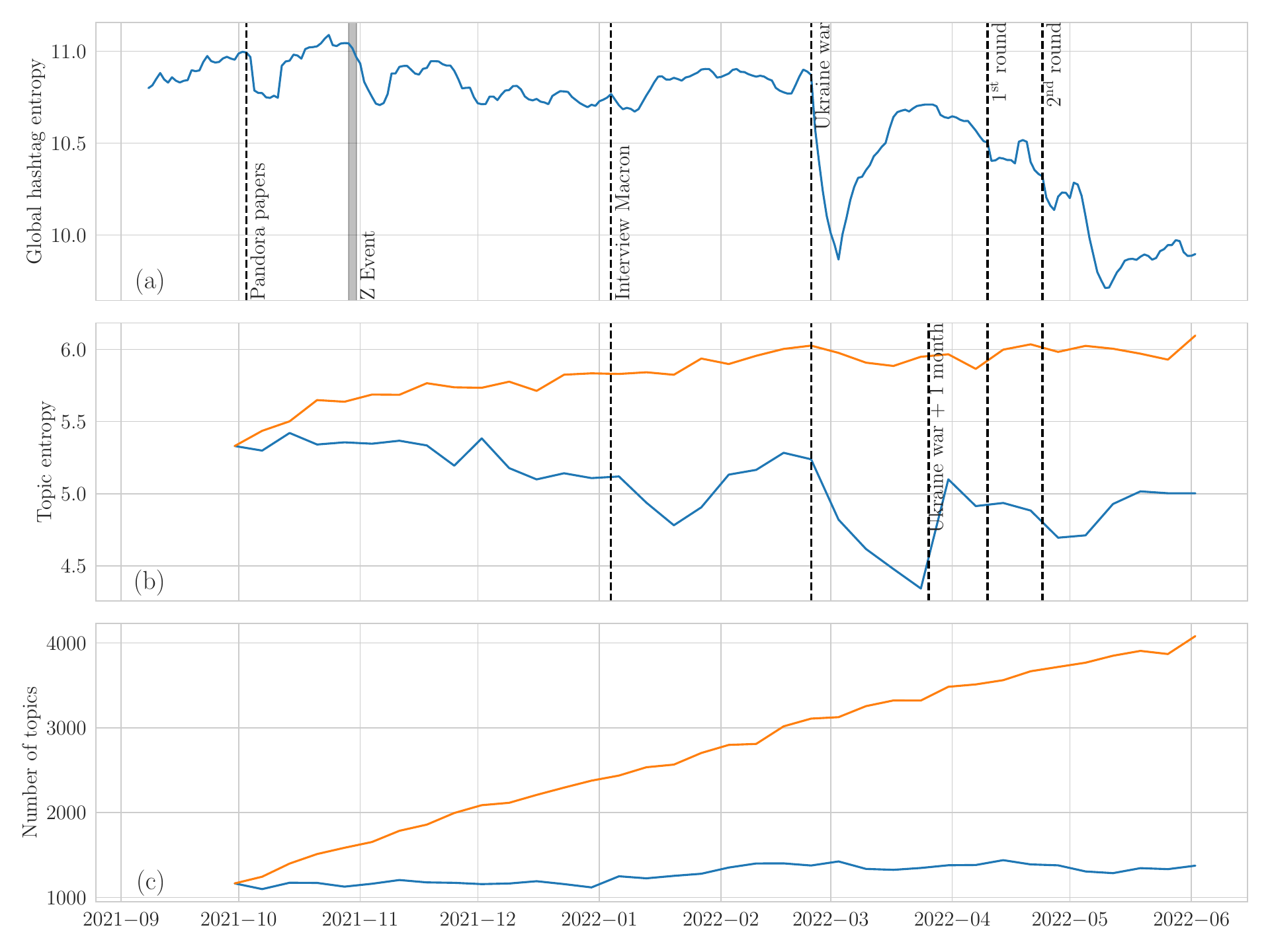}
	\caption{Hasthags' and Topics' entropy evolution. (a): Time evolution of the global hashtags' entropy. Notice that in this case the method of constructing the semantic network is irrelevant. (b): Comparison of the time evolution of the topics' entropy, computed using the rolling window and aggregated co-occurrence networks. The vertical lines mark the dates of occurrence of important specific events. (c): Comparative evolution of the number of topics between the two methods. For (b) and (c) the orange and blue curves correspond to the growing aggregated and rolling windows approaches, respectively. }
 \label{fig_entropy}
\end{figure}

Fig.~\ref{fig_entropy} shows the hashtag and topic entropy, as well as \lh{the evolution }of the number of topics \lh{ for the two methods of construction of the semantic network. }

A high entropy indicates a large variety of hashtag usage, 
a low entropy indicates that fewer hashtags are being used, i.e.,  attention is concentrated on  few of them. 
The last situation is \lh{characteristic of} certain events, such as the Ukraine war, \lh{clearly visible} in the hashtag entropy, where the entropy drops to a local minimum. 
\lh{Vertical bars in Fig.~\ref{fig_entropy} indicate other important events that led to a sudden decrease of hashtag entropy, like} the Pandora Papers (the International Consortium of Investigation Journalists revealed that hundreds of world political leaders, celebrities and business leaders held between 5.6 and 32 trillion US dollars in offshore accounts, in the so-called ``tax paradise" countries) and the ZEvent (online charity event that gathered the most popular French content creators over the course of three days and on this 5th edition raised 10 million euros for the NGO Action Against Hunger). 
The French presidential elections also attracted some attention, but  less than the aforementioned. 

The topic entropy shows different results according to the model followed, although it is in general less sensitive than the hashtag  entropy.  

\lh{The lack of sensitivity shown by the entropy in the case of the aggregated network  suggests  the
new hashtags added weekly,  simply attach to existing topics, and even major events such as the Ukraine war are not enough to impact the consolidated structures of the network. On the contrary, for the rolling window update, the entropy shows some of the  features of the hashtag entropy. This comes from its ability to discard old information. The network  ``forgets" at a rate given by the width and step of the rolling window. This is clearly visible in Fig~\ref{fig_entropy} for the sudden drop at  the breaking of the Ukraine war: the topics' entropy with the rolling window method returns to almost its pre-war level exactly after one month. The issue is not as ``hot" and is less present in the users’ mind, several hashtags that constituted the topic related to the burst of the conflict are are not used anymore and the network ``forgets" them once they are past its
window (here after a month).}

\subsection{Similarities}

\begin{figure}[t!]
	\centering
 \includegraphics[width=\linewidth]{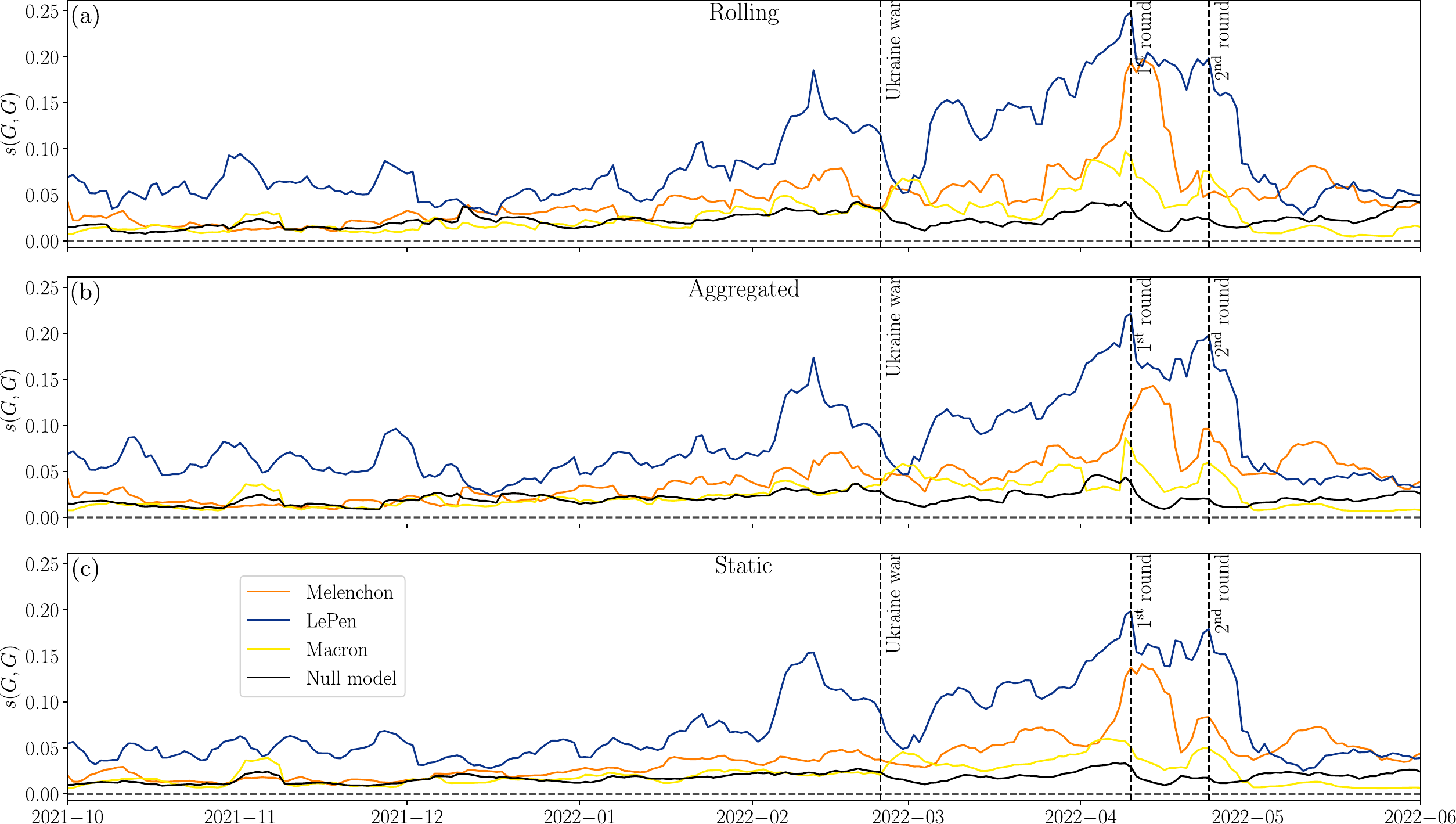}
	\caption{Time evolution of the self-similarity of the three candidates having obtained the largest amount of votes in the election \textit{Macron, Le Pen} et \textit{Mélenchon} along with that of the null model, for comparison. Vertical lines signal the begining of the Ukranian war and the date of the two rounds of the 2022, French presidential election. (a) Measurements corresponding to the \textit{rolling window network method}. (b) Measurements corresponding to the \textit{aggregated growing network}  method. (c) Measurements corresponding to the \textit{static network} method, for comparison}\label{self_sim}
\end{figure}

\lh{In this section we show how the cohesion among different groups of supporters evolves with time when measured by the different semantic networks, built as described in Section~\ref{similarities}. We also show the evolution in time of the cross-similarity between a  group of supporters of a political candidate and a group of users who hold a preferred news medium.} 

\lh{It is interesting to keep in mind what differs in the computation of similarity using the information issued from different networks: while the hashtags used by a given user in a given day are fixed, the same hashtag may refer to different topics, according to the semantic network we are considering (rolling windows, growing aggregated or static). The similarity then reveals the differences, if any, related to the  method chosen to build the co-occurrence networks.}

Fig.~\ref{self_sim} shows the \lh{time evolution of the } self-similarities of 
\lh{the groups supporting each of the three}  main candidates: 
\textit{ M\'elenchon, Le Pen}  and \textit{Macron}  \lh{(representing the left, the extreme right and the center-rigth parties,  respectively; the latter being in power before the election)}. 
\lh{For the sake of comparison, we have added an extra similarity curve  that represents} a \emph{null model} group computed from 5,000 randomly selected active and ``in France'' users.

 \lh{As the self-similarity is an indicator of the cohesion within the group,  we see that for each of them it is globally positive in the three panels, and very low for the group constituted of random users, as expected.  All the panels show that the self-similarity of the group of supporters of \textit{Macron} is  lower than the others and very near the null model, which can be explained by the fact that, as he was the incumbent president,  his audience was   more diverse than for the other candidates. }

 \lh{The self similarity also captures that the group of extreme right (\textit{Le Pen}) appears much more cohesive than the others. The most relevant characteristics, like the peaks showing an increase of the cohesion, appear  at similar dates for the three methods of computing the semantic networks. }

 \lh{However, a comparison of the similarity curves of the same political group obtained by  different semantic networks reveals that they do not provide exactly the same information. For example, if we consider the self similarity of the supporters of \textit{M\'elenchon}, one sees that around the second round of the presidential election, the results of the aggregated and static network are similar but differ from those of the rolling window case: the small peak observed in the first two, is absent from the latter. 
    Far from being an error, this difference reveals that this method is capturing the internal turmoil that caught the supporters of this political option, which could not access the second round, when they had to decide the action to take next. As they are historically opposed to the extreme right but in complete disagreement with the incumbent president, the dissensions during this week were intense. This fact can only be captured by the rolling windows method, which gives relatively more importance to the new hashtags entering at each weekly update than the other two methods, where they are lost in a much larger network. }

\begin{figure}[t!]
	\centering
 \includegraphics[width=\linewidth]{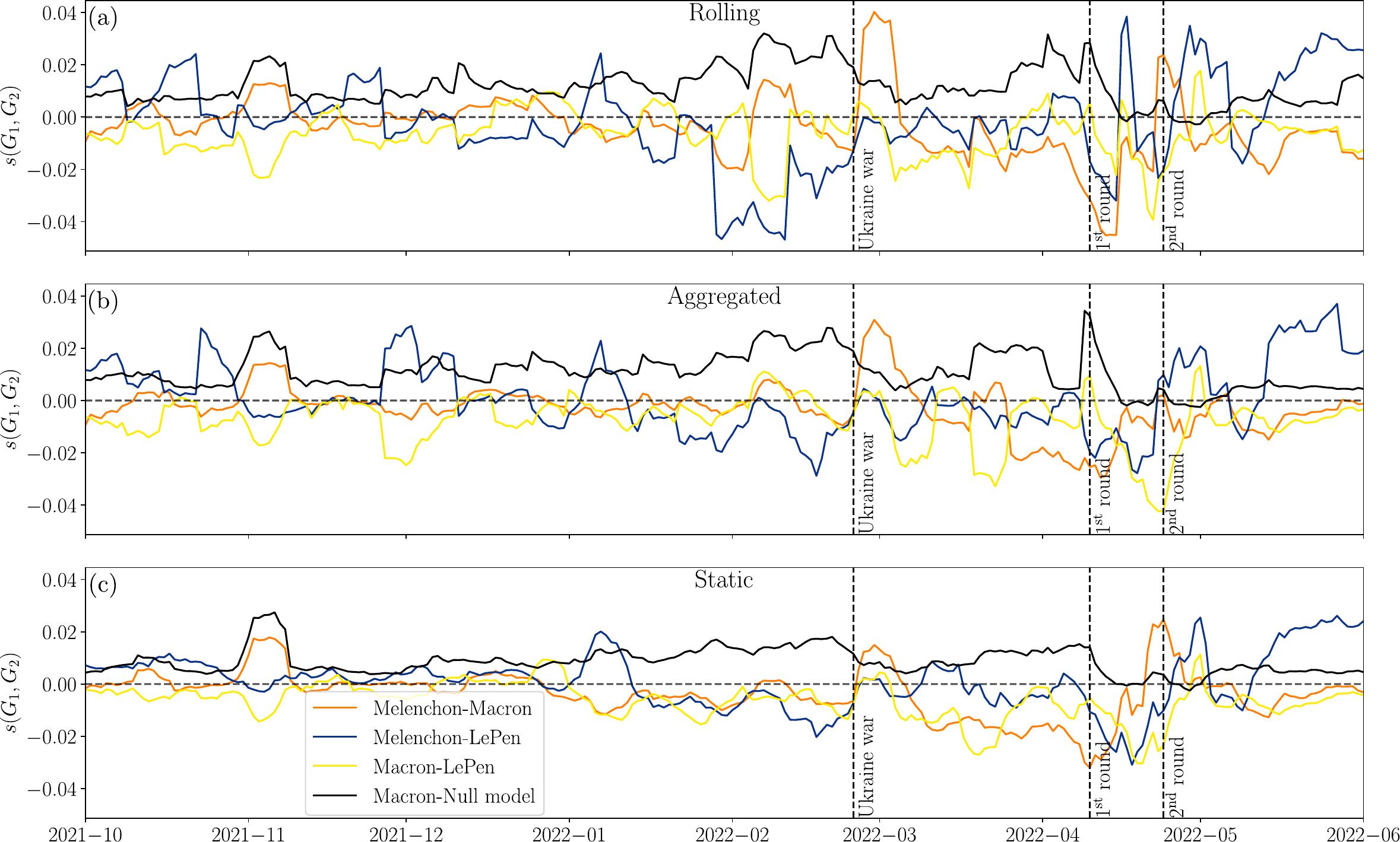}
	\caption{Time evolution of the cross-similarity between the different pairs formed by the three candidates having obtained the largest amount of votes in the election \textit{Macron, Le Pen} et \textit{Mélenchon}. For comparison the cross-similarity between the winner \textit{Macron} with  the null model, is also shown. Vertical lines signal the begining of the Ukranian war and the date of the two rounds of the 2022 French presidential election. (a) Measurements corresponding to the \textit{rolling window network} method. (b) Measurements corresponding to the \textit{aggregated growing network}  method.  (c) Measurements corresponding to the \textit{static network} method, for comparison.}\label{cross_sim_candidates}
\end{figure}

Fig.~\ref{cross_sim_candidates} shows the comparison of the cross-similarities between \lh{groups supporting different candidates.}

\lh{Here we have a confirmation  that the supporters of \textit{Macron}, the incumbent president, behave like the null-model sample: the cross-similarity of his supporters with a random sample is higher than the cross-similarity with the other groups, regardless the method. The lower comparative sensitivity of the data obtained by the static network suggests that this method misses information. }

\lh{Focusing on the dynamical networks, besides the negative global trend for the cross-similarities between the supporters of the extreme right (\textit{Le Pen} ) and those of the other two parties, which is expected, the peaks in these curves reveal that at some particular times, the  considered pair of groups are activating the same topics more frequently than the general population. One interesting case is the burst out of the Ukranian war, where clearly \textit{Le Pen} supporters were not activating the same topics as the other two groups.  
Although there is a global coherent trend in the curves issued from the two dynamical networks (panels (a) and (b) ), the peaks are not exactly the same. Again this shows that keeping all the history in the aggregated network, may lead to wash out the detection of events of short duration, but nonetheless, significant. A good example is the very sharp peak between the two rounds of the presidential election in the cross-similarity curve (Fig.~\ref{cross_sim_candidates} ) between \textit{M\'elenchon} and \textit{Le Pen} supporters that appears in the rolling windows case (blue curve in top panel)  but is absent from the aggregated one. This period of two weeks, corresponds to the turmoil described above, where \textit{Mélenchon} supporters participated in online discussions, that activated topics criticizing the incumbent president, thus making them, for a very short period similar to the group of the pretending candidate. This is not  captured by the aggregated --long memory-- method, because the new hashtags introduced by this discussion are attached to older topics, which have been forgotten in the rolling windows method.}

\begin{figure}[]
	\centering
 \includegraphics[width=\linewidth]{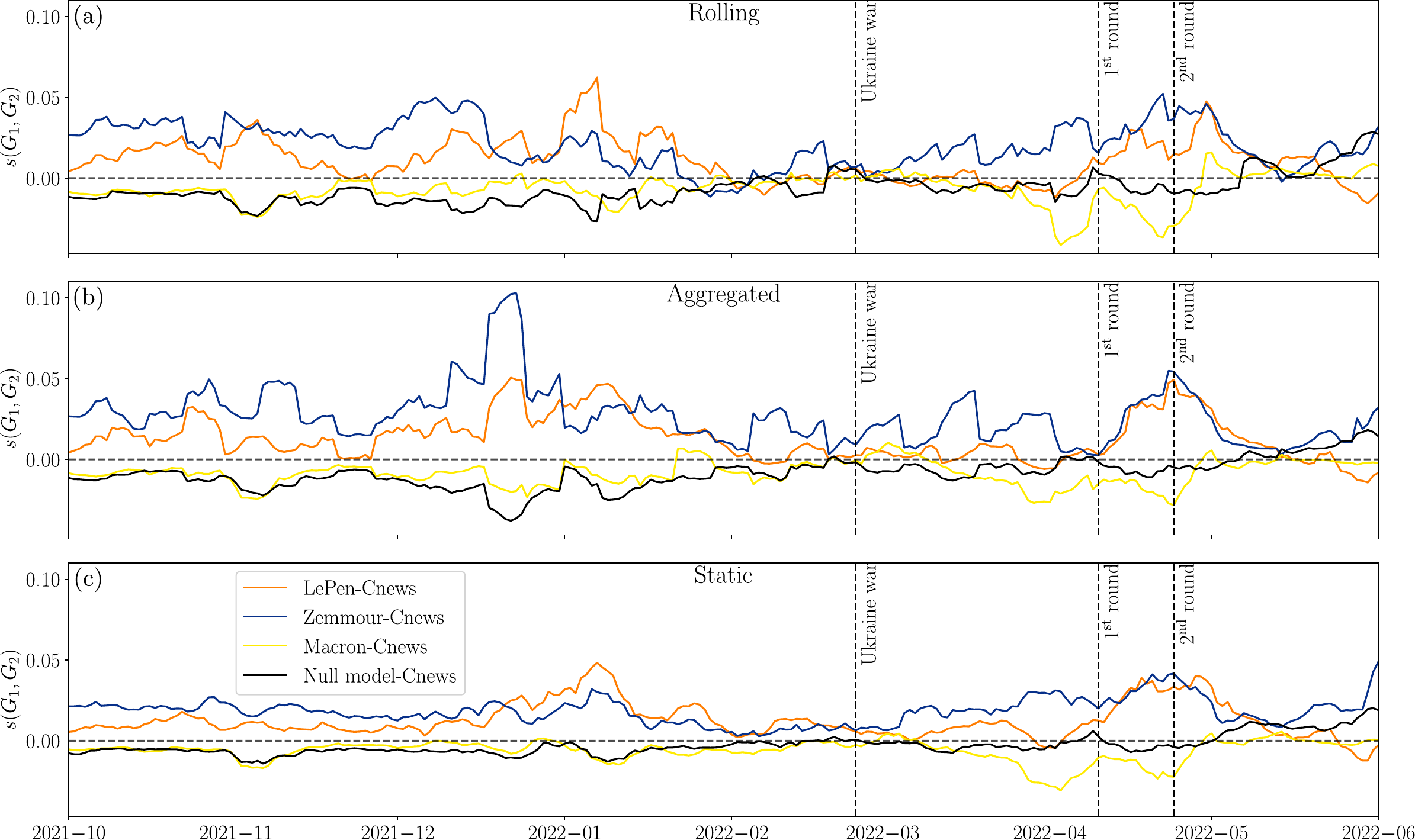}
	\caption{Time evolution of the cross-similarity between the continuous news TV chain \textit{CNews} and the three candidates having obtained the largest scores \textit{Macron, Le Pen} and \textit{M\'elenchon}, for comparison the cross-similarity of \textit{CNews} and the null model is also shown. (a) Measurements corresponding to the \textit{rolling window network} method. (b) Measurements corresponding to the \textit{aggregated growing network}  method.  (c) Measurements corresponding to the \textit{static network} method, for comparison.}\label{cross_sim_medias_candidates}	
\end{figure}

\lh{In Fig.~\ref{cross_sim_medias_candidates} we plot the cross-similarity between the group of users having \textit{CNews} (a continuous news TV chain, classed as favourable to the extreme right) as preferred media and the group of supporters of two extreme right  candidates --\textit{Le Pen }and \textit{Zemmour}-- and for comparison, the group of supporters of the incumbent. All the methods of building the network detect the positive similarity between those users that get their information preferentially from \textit{CNews} and the two extreme right candidates. Notice  that  the calculation does not include those users at the intersection of two groups so as to avoid a spurious signal. Interestingly, though  positive for both extreme right supporters, the cross-similarity with users preferring \textit{CNews}  is higher with  \textit{Zemmour} supporters. This is less surprising considering that this candidate was a former political journalist in the \textit{CNews}  chain before running for president.}

An interesting observation on the aggregated model is the peak observed at the end of the year for the \textit{Zemmour} case, which is not found on the rolling model. 
This \lh{is an interesting example as it shows that }  one topic was activated on the aggregated model, \lh{but the same hashtag usage does not activate a topic in the rolling model.} 
An investigation of the \lh{structure and composition of the } topics of the three networks at that \lh{precise} time revealed that the aggregated network \lh{concentrates the hashtags used by  \textit{CNews} and \textit{Zemmour} groups in a single topic, which is responsible of the peak in the cross-similarity. }
In contrast, on the rolling model, \lh{the same set of hashtags}, \lh{are distributed among different topics, so their activation does not lead to a clear peak},  while on the static model these hashtags were \lh{concentrated} in few topics but diluted due to the size of these topics (i.e., a low activation).

The \textit{CNews-Macron }cross similarity is very close to the \textit{CNews-}null\_model one, and most of the time with negative values. 

\begin{figure}[t!]
	\centering
 \includegraphics[width=\linewidth]{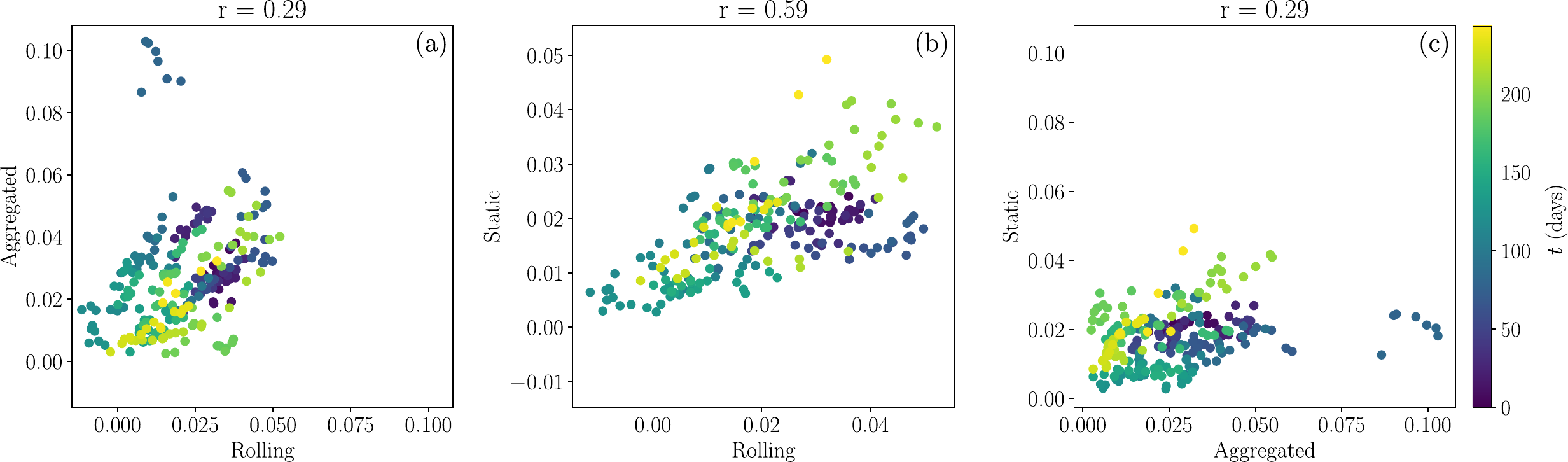}
	\caption{Scatter plots representation of the time series of the cross-similarity between \textit{Zemmour-CNews}, grouped by pair of methods. This representation reveals the correlation between each pair of methods. The time stamp of the measurement of the cross-similarity is given by the color map. The Pearson coefficient is indicated in the top of each panel. (a):  Comparison of the cross-similarity measured by the \textit{growing aggregated and rolling window} methods. (b): Comparison of the cross-similarity measured by the \textit{static and rolling window} methods. (c): Comparison of the cross-similarity measured by the \textit{static and growing aggregated} methods. }\label{scatter-plots-sim}
\end{figure}

\lh{In order to have a global way to characterize the analogies or differences that we obtain when using different semantic networks models, we compute the scatter plot of the time series of each cross-similarity measured by two different methods.}
 Fig.~\ref{scatter-plots-sim} shows the scatter \lh{plots of the time series of the cross-similarity between the \textit{Zemmour} and \textit{CNews} groups, from Fig.~\ref{cross_sim_medias_candidates}; each panel shows a different pair of  networks used in the calculation and the time-stamp of each point is given in a color scale. In general, a rough global positive correlation is observed, and in the panels that include the aggregated network, the  outliers (in light blue), correspond to the already discussed peak around the end of 2021, only present in the aggregated network model.} 

\begin{figure}[h!]
	\centering
    \includegraphics[width=0.82\linewidth]{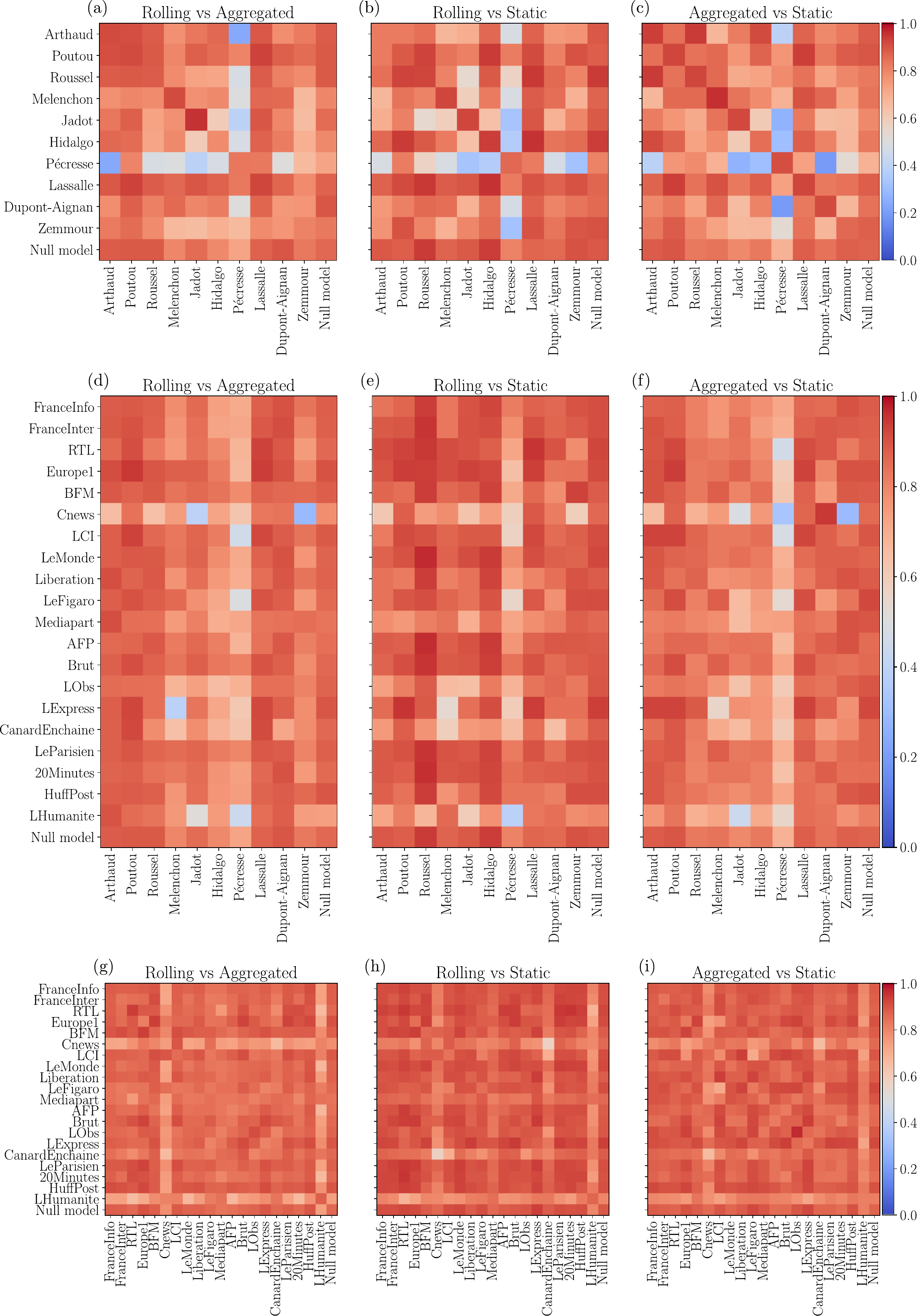}
	\caption{Color-coded matrices of Pearson coefficients comparing the correlation of similarities obtained by different pairs of models (\textit{static, aggregated, rolling}) for each pair of supporters' group of candidates and media. Top row: self and cross similarities among groups of supporters of different candidates (a) \textit{rolling window} vs. \textit{growing aggregated} method, (b) \textit{rolling window } vs. \textit{static} method, (c) \textit{growing aggregated} vs. \textit{static} method. Middle row: cross similarities among groups of supporters of different candidates and users having a preferred media. (d) \textit{rolling window } vs. \textit{growing aggregated} method, (e) \textit{rolling window } vs. \textit{static} method, (f) \textit{growing aggregated} vs. \textit{static} method. Bottom panel: self and cross similarities among groups of users having a preferred media. (g) \textit{rolling window } vs. \textit{growing aggregated} method, (h) \textit{rolling window } vs. \textit{static} method, (i) \textit{growing aggregated} vs. \textit{static} method}.\label{matrices_sim}
\end{figure}

Following this procedure for each cross-similarity and each pair of network models, we can compute in each case the Pearson coefficient that gives the correlation of results issued from the two considered methods of network construction. The results are depicted in the correlation matrices of Fig.~\ref{matrices_sim}. In each panel we compare the Pearson coefficient of all the results obtained by a pair of network models. The first line of panels gives all the possible cross/self-similarities among groups of supporters of candidates, the second line describes the cross-similarities between supporters of candidates and  media, and the bottom line  gives all the possible cross/self-similarities among groups of users having a preferred media. 
The first general observation is the consistency of the methods: all the correlations are positive, and mostly quite high.

\lh{In some cases}, it is possible to identify \lh{a few }instances where the correlation between the results obtained by different network methods is lower than $r=0.5$ \lh{(represented by blue pixels). This can be observed for example, in the first row that involves cross/self-similarities between the supporters of the candidates, only for the supporters of \textit{Pecresse} candidate. As this discrepancy between methods is present for all the pairs and only for this group, we suspect that it is associated to the group and not to the method used for computing the networks. In the second row, besides the cited case,  and the case of the groups of \textit{Zemmour} and \textit{CNews}, discussed above, we also find a very low Pearson coefficient  for the cross-similarities between groups of candidates' supporters and groups having a preferred media (intermediate line). These lower correlation values are found for  the candidate \textit{Jadot} and the medium \textit{CNews}, in the two panels that consider the aggregated network. This discrepancy in the methods is similar to the already explained for the outliers observed in the \textit{Zemmour-CNews} cross-similarity (see Fig.~\ref{scatter-plots-sim}), there  is a difference in hashtag-topic association in the rolling windows and aggregated network. }

\section{Discussion}\label{sec4}

In this work, we present an empirical study based on a large data base collected from Twitter (before it became X), during the French presidential elections of 2022.
We aim to understand the distribution of the attention of different groups of users over the topics of discussion in the platform. In particular, we want to detect the potential correlations in the dynamics of attention between groups of supporters of the different candidates and groups of users with a preferred media. 
We propose a methodology that allows us to develop, as far as possible, a \textit{real time} dynamical study. To do so, we work with two alternative methods to build dynamical semantic networks (co-occurrence hashtag networks): the \textit{growing aggregated network}, which keeps track of all the hashtags produced since the beginning of the study,  and the \textit{rolling window network}, which has a bounded memory, and ``forgets" hashtags that have not been used within a month. Consequently, we have two different descriptions of the topics treated in the platform that we use to measure the similarity between the groups.

Compared with previous works, which use a static network aggregated throughout the period~\cite{mussi_reyero_evolution_2021,schawe_understanding_2023}, the methods developed here produce more reliable results, not only in the sense that they do not use information \textit{from the future}, but also because we show that dynamical methods (both rolling window and growing aggregated network) are in general more sensitive. 
 
Our results show that it is possible to detect the synchronization of the attention of groups of supporters of a candidate with that of users seeming to prefer one particular media. It is very important to stress that this can be done in a completely anonymous and agnostic way: we do not follow one particular user, but a group characterized but their public expression;  we do not define the topics of discussion \textit{a priori}: they emerge from the community structure of the semantic network.

Moreover, using simple metrics like the hashtag entropy, it is possible to reveal the internal state of conversations among the supporters of a candidate. For example, we observe that the vocabulary used by \textit{Zemmour}'s supporters remained largely unaffected by the burst of the Ukraine war, as compared with the vocabulary used by supporters of other candidates whose entropy show a sudden decrease, thus revealing a more concentrated discussion sphere. Interestingly, the hashtag entropy of the same group is in general the lowest throughout the whole period. This is an unexpected result because as the entropy is extensive in the number of hashtags, its value  is  expected to be low for a group that uses  few hashtags, which can naturally happen for a small group of supporters. This is not the case here, Fig.~\ref{stats_candidates_media} shows that \textit{Zemmour}'s supporters constitute one of the largest groups online, so the low value of their entropy reflects a choice of those users who restrict their numerous interventions to a limited vocabulary  (see entropy curves discriminated by candidate in the Appendix).

The observed synchronization of interests between users having as preferred media \textit{CNews} --a TV chain providing 24-hour news coverage-- and those supporting the far-right candidate \textit{Zemmour} is extremely interesting, although not surprising. 
There is a long-lasting controversy in France about the editorial line of this TV chain,  which has recently been the object of an official investigation from a commission of the Parliament~\cite{commission_parlamentaire_TNT}. The broadcasting choices of the chain and their compliance with law  have also been the subject of extensive studies carried on by social scientists  who have followed hours of programs trying to qualify the type of speech and to quantify its frequency~\cite{c_secail}. However, they are often disregarded by the defenders of the chain, arguing that the results of these works are based on a selected sample of few hours, that cannot be taken as representative of the global content that is diffused. Our results come to complement these works: we analyse \textit{all} the discussions held during a long period  by those users who have \textit{CNews} as a preferred media,  in comparison with the speech of the supporters of all  the candidates, and we automatically detect the same features observed in qualitative studies.  Certainly, they can argue that this is not what has actually been diffused, but the conversations on Twitter of their followers, nevertheless the chain \textit{selects} what is published on the platform that is then retweeted by their supporters. 

 The two dynamical methods of construction of the semantic network developed here are complementary: while the growing aggregated network model detects the possibility of reactivation of topics by allowing  new incoming hashtags to attach to old ones that have been less used recently, the rolling window model is more sensitive to important, though short, events (as in the discussions during the two weeks in between the two rounds of the presidential elections, for instance). 

 Strictly speaking,  the study presented here is a \textit{quasi} real time approach because of the method we used  to  define the supporters of the candidates or the users having a preferred media.  For simplicity, we choose to  tag users as supporters of a candidate (or having a preferred medium) when more than $75\%$ of their total retweets of candidates (or media) correspond to a single candidate (or medium).  Specifically, we detect ``supporters"   by using information about their retweet behaviour over the whole period, which is inconsistent with a true ``real-time" approach. However, we kept this choice to improve the statistics of the``small candidates". To check this point,  we have  performed an update of the candidates each week, in synchrony with the update of the semantic networks. Our results (see Appendix \ref{secA3}) show that this is not a good solution because it captures spurious  effects that  may blur the results. We observe candidates gaining supporters and then loosing them immediately. An inspection of these particular events, reveals that they correspond to a buzz created by the candidate or media that suddenly boosts the retweets of their messages, an effect that disappears soon after.  So including these users as supporters of the candidate or a medium would be misleading.  A way to avoid using information from the future, would be to determine all the supporters only at the initial month. In an electoral period, where one expects a growing attention of the population as the elections approach, this method risks to  let out of the analysis the newcomers that integrate the discussion as the campaign advances.  There is a  compromise to establish between the weekly update and the static detection of supporters, one possibility would be to recompute supporters at  longer periods in order to improve statistics while diminishing the buzz effect. Nevertheless, this way to detect supporters is independent of the rest of the proposed method, and could be performed also by incorporating external information. In any case, the results shown here, which concern the most important candidates and media, are the less susceptible to suffer from the choice in the determination of supporters.

\section{Conclusion}\label{sec5}

The dynamical study of discussions in Twitter among the  supporters of different candidates to the French presidential election of 2022, and the followers of different types of media, is able to automatically detect  the synchronization  of interest of different groups, around particular topics at particular dates.

We propose two complementary methods to obtain dynamic semantic networks, each of them having a particular interest: while the \textit{growing aggregated network }allows to detect the reactivation of old topics, the \textit{rolling windows network} is more sensitive to   topics  that, though important, may appear suddenly and have a short life.  Each of the methods represents the discussion landscape from a distinct perspective, so that  instead of asking which one to choose,  one should  rather  keep them both because at a fairly low cost in term of computation and storing space, useful information may be obtained by comparing their results. 

Our results  confirm and quantify qualitative observations reported in previous works~\cite{c_secail,commission_parlamentaire_TNT}, using an agnostic and automatic procedure that allows for data treatment at a larger scale.   In this respect,  this work constitutes a step forward in the development of methods aimed at measuring equity in the treatment of information, a requirement that the French law imposes to broadcasters using public spectrum bands.

\backmatter

\bmhead{Acknowledgments}
R.P. and L.H. acknowledge the financial support of Labex MME-DII,  (Grant No. ANR reference 11-LABEX-0023). J.I.A.H. and M.G.B.  acknowledge the financial support of UBACyT 2023 20020220100053BA. 

\section*{Declarations}

\begin{itemize}

\item The authors declare no conflict or competing interests
\item Authors' contributions: 
(R.P.) performed the data collection, curation and coded all analysis  and visualization tools, (L.H.) proposed research subject, supervised development, analysed and discussed results, wrote the manuscript and secured financial support, (J.I.A.H.) analysed and discussed results, wrote the manuscript, (M.G.B.) assisted with data collection, analyzed and discussed results, contributed to the manuscript, (D.K.) analyzed and discussed results, ALL checked final version of the manuscript.
\end{itemize}

\clearpage

\begin{appendices}

\section{Supplementary Information }\label{secA}
\subsection{Distribution of the number of hashtags per tweet}
\begin{figure}[h]
	\centering
 \includegraphics[width=.6\linewidth]{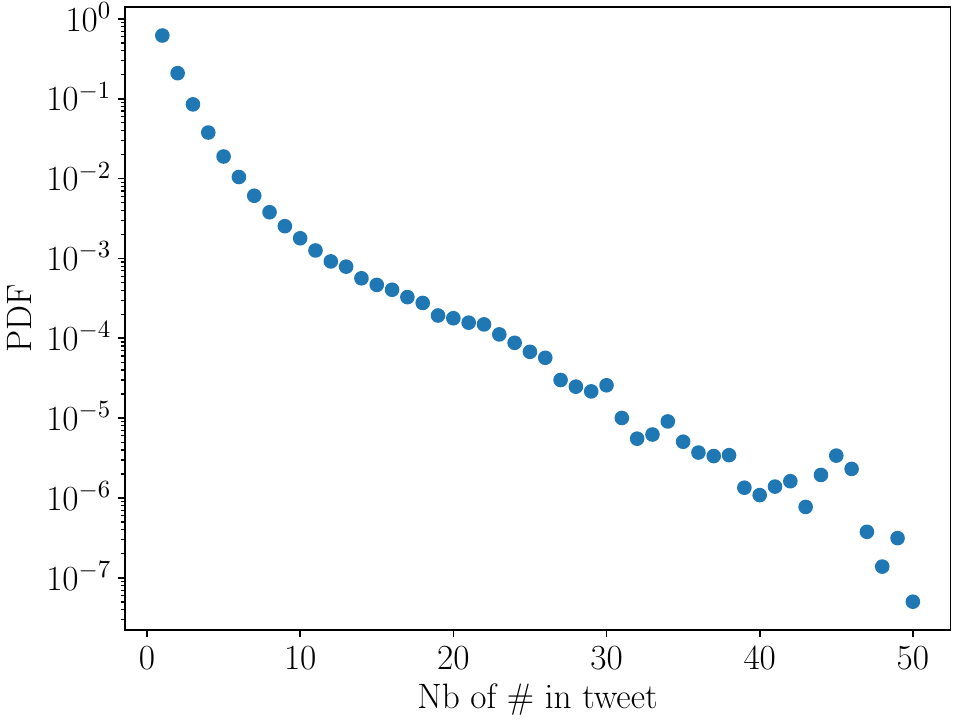}
	\caption{Distribution of the number of hashtags per tweet. No filtering has been applied here, and all tweets containing hashtag(s) are considered.}\label{distrib_hashtags_per_tweet}
\end{figure}

Fig.~\ref{distrib_hashtags_per_tweet} shows the distribution of the number of hashtags per tweet (ignoring tweets with no hashtags), computed over the entire collection. It is unclear whether the cut-off at 50 is due to some hard-coded limit on the maximum number of hashtags in a tweet, or if it is simply due to the fact that one cannot put many more intelligible hashtags in 280 characters.

\begin{figure}[t!]
	\centering
 \includegraphics[width=\linewidth]{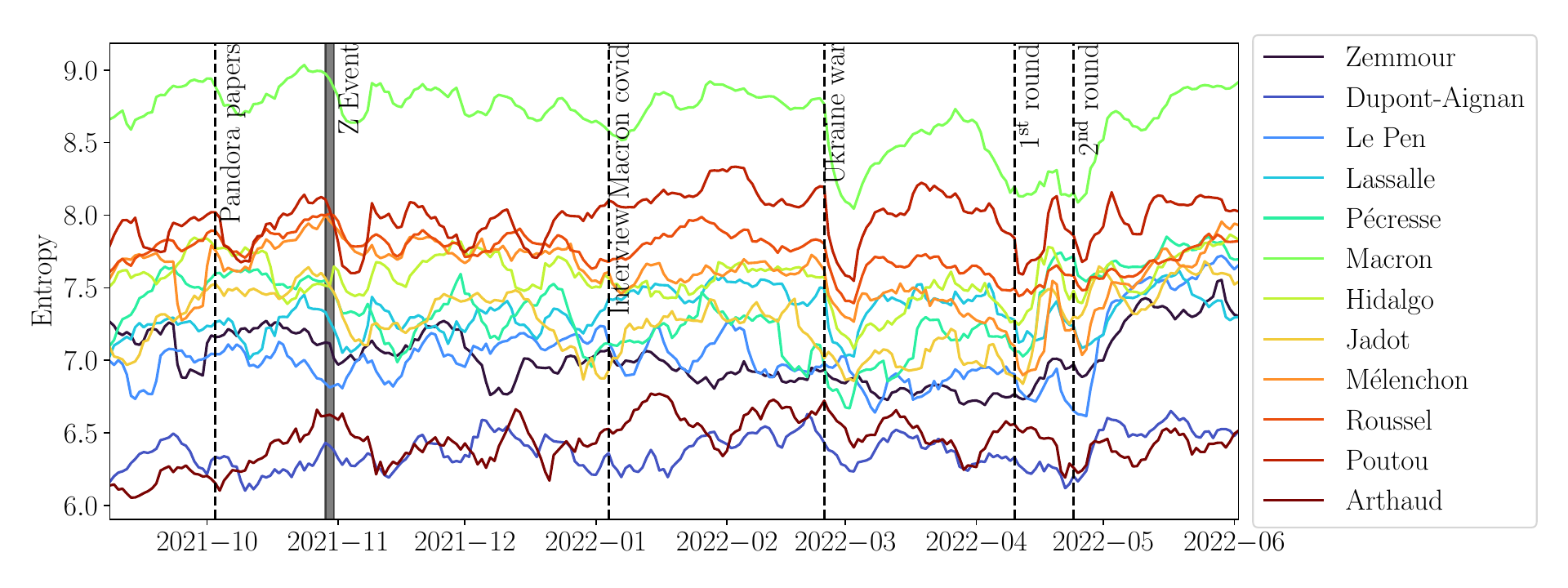}
	\caption{Evolution of the hashtag entropy disclosed by usage of supporters of different candidate. The vertical lines mark dates of events that attracted attention in the online discussions in France.}\label{hashtag_entropy_candidate}
\end{figure}
\subsection{Hashtags entropy discriminated by candidates' supporters}
Fig.~\ref{hashtag_entropy_candidate} shows the hashtag entropy for the supporters of each of the candidates. As the entropy is an extensive quantity, the curves are expected to be ordered in such a way that the lowest values correspond to groups with the smallest number of supporters. Interestingly, 
this is not the case here, where the curve corresponding to \textit{Zemmour} supporters which are not the smallest group (see Fig.~\ref{stats_candidates_media} in the main text), has comparatively very low entropy. This reveals that the vocabulary (thus the topics) used by this group is only a small fraction of the one used in the general population.

\subsection{Updating supporters with the frequency of the semantic network}\label{secA3}
\begin{landscape}
\begin{figure}[h]
	\centering
 \includegraphics[width=\linewidth]{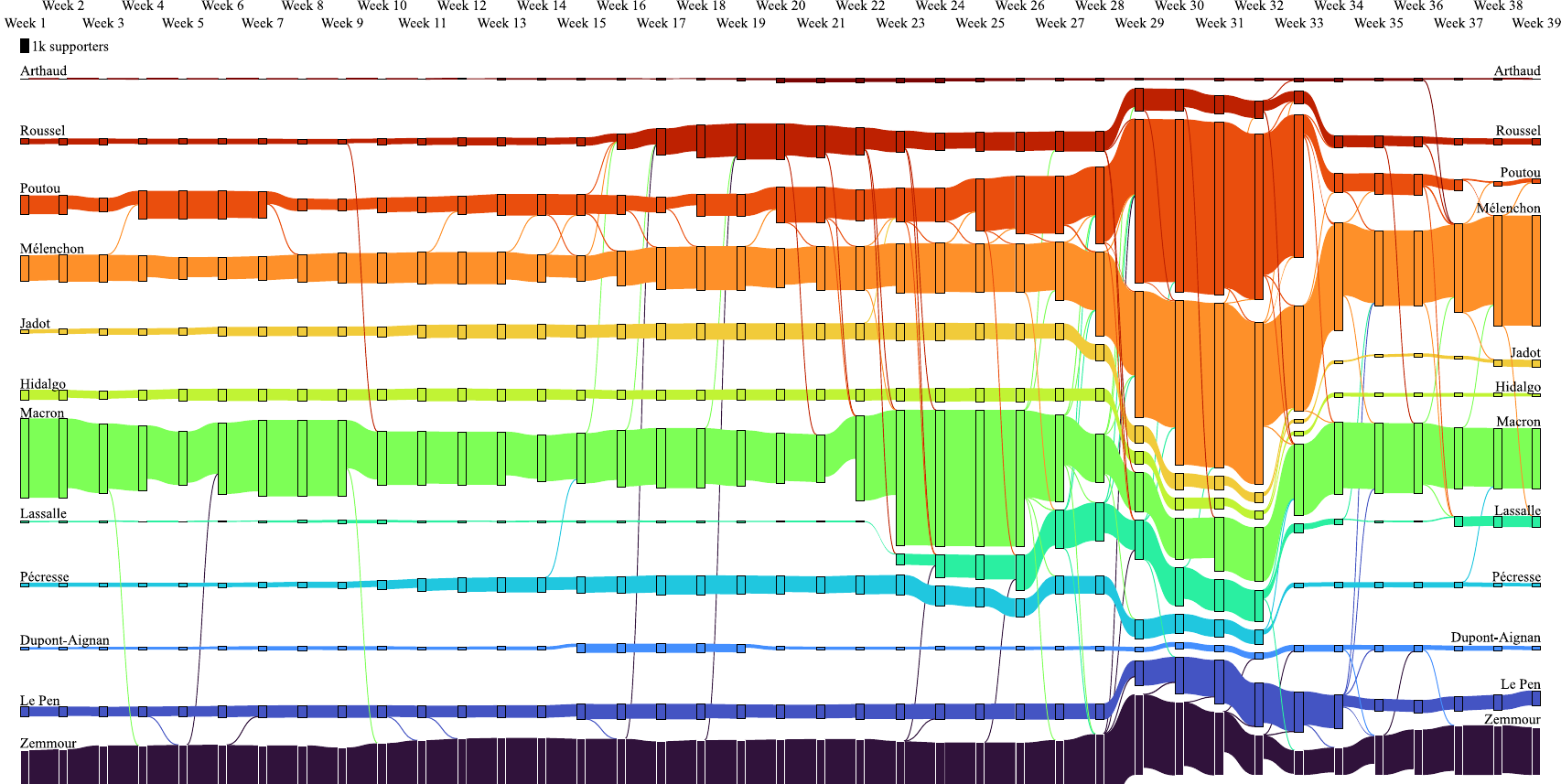}
 \caption{The flowchart shows the evolution of the supporters of the candidates as a function of time during the campaign . For the sake of legibility, transfers between candidates are only displayed if they involve more than 5 supporters. A user is considered a supporter of a candidate if 75\% of their retweets correspond to the candidate. The presidential elections took place on 10 April 2022 (1st round) and 24 April 2022 (2nd round), which corresponds to Week 29 and 30.}\label{flowchart}
\end{figure}
\end{landscape}

We have  performed an update of the candidates each week, in synchrony with the update of the semantic networks. Fig.~\ref{flowchart} shows that this is not a good solution because it captures spurious effects that may blur the results. We observe candidates gaining supporters and then losing them immediately (see the example of candidate \textit{Poutou} representing a small extreme left party) . An inspection of these particular events reveals that they correspond to a buzz created by the candidate or media that suddenly boosts the retweets of their messages, an effect that disappears soon after.

\newpage

\newpage

\end{appendices}

\bibliography{zotero_export}

\end{document}